\documentclass[sigconf]{acmart}

\usepackage{color}
\usepackage{hyperref}

\usepackage{booktabs} 

\usepackage{etoolbox}
\makeatletter
\patchcmd\@combinedblfloats{\box\@outputbox}{\unvbox\@outputbox}{}{%
   \errmessage{\noexpand\@combinedblfloats could not be patched}%
}%
 \makeatother

\setcopyright{rightsretained}

\acmDOI{10.475/123_4}

\acmISBN{123-4567-24-567/08/06}

\copyrightyear{2018} 
\acmConference[RecSys '18]{ACM Recommender Systems conference}{2nd-7th October 2018}{Vancouver, Canada} 	
\acmYear{2018}

\acmArticle{4}
\acmPrice{15.00}

\begin{document}
\title{Action-conditional Sequence Modeling for Recommendation}

\author{Elena Smirnova}
\affiliation{%
  \institution{Criteo Research}  
  \city{Paris}   
}
\email{e.smirnova@criteo.com}

\begin{abstract}
In many online applications interactions between a user and a web-service are organized in a sequential way, e.g., user browsing an e-commerce website. In this setting, recommendation system acts throughout user navigation by showing items. Previous works have addressed this recommendation setup through the task of predicting the next item user will interact with. In particular, Recurrent Neural Networks (RNNs) has been shown to achieve substantial improvements over collaborative filtering baselines. In this paper, we consider interactions triggered by the recommendations of deployed recommender system in addition to browsing behavior. Indeed, it is reported that in online services interactions with recommendations represent up to 30\% of total interactions. Moreover, in practice, recommender system can greatly influence user behavior by promoting specific items. In this paper, we extend the RNN modeling framework by taking into account user interaction with recommended items. We propose and evaluate RNN architectures that consist of the recommendation action module and the state-action fusion module. Using real-world large-scale datasets we demonstrate improved performance on the next item prediction task compared to the baselines. 
\end{abstract}

%
%



\maketitle

\section{Introduction}
\label{sec:intro}
Many applications of recommender system (RS) naturally involve sequential interaction between user and the system, e.g., user navigating an e-commerce website to satisfy her shopping needs or user reading news online. Motivated by these applications, we consider the following setup.


User is navigating across the website which can be seen a sequence of timestamped interactions. Each interaction involves one or more items and is associated with a predefined type (e.g., in e-commerce, item-view, item-purchase, add-to-cart). Such form of data is typically available in the form of interaction logs \cite{yoochoose:dataset,xing:dataset}.

In addition to navigations on the website, we consider the production RS that is deployed on the website. This recommendation system acts by presenting a list of recommended items throughout user navigation \cite{jugovac2017interacting}. The navigational data of the user including his interactions with recommendations is used to learn a new RS. 

The framework of learning from logged data is frequent in RS research\cite{yoochoose:dataset,xing:dataset,dataset:30music}. Typically, datasets only contain the navigational data, but not the recommendations of the deployed RS~\cite{yoochoose:dataset,dataset:30music}. Differently, in this work we are interested in modeling the user behavior including the interaction with the RS.

Indeed, according to multiple studies~\cite{grau2009personalized,mulpuru2006you,sharma2013pairwise}, up to 30\% of total interactions is coming from interactions with recommendations. Moreover, in practice, RS often act to achieve companies' business goals, e.g., increase basket size or promote specific products~\cite{jannach2016recommendations,jannach2017price,jannach2017session} and thus, influence user behavior~\cite{kamehkhosh2018automated,said2013month}.

In this work, we account for the RS impact by extending previous studies on sequential modeling in recommendation~\cite{sequence_survey}, in particular, session-based recommendation~\cite{rnn:balazs}.
Similarly to these works, we will focus on the task of predicting the next item user will interact with. On this task, Recurrent Neural Networks (RNNs) have shown to substantially outperform the collaborative filtering baselines. Recently, multiple extensions have been proposed including item content modeling~\cite{feature-rnn:balazs}, context modeling~\cite{smirnova2017contextual,Twardowski:2016:MCI:2959100.2959162} and cross-session modeling~\cite{hrnn}. We propose a new extension to the RNN architecture that takes into account the interactions triggered by the recommendations of production RS. 

Previously, the interaction with recommendations have been studied separately in the context of click prediction tasks~\cite{xing:dataset,kasandr_dataset,yoochoose:dataset,outbrain:dataset}. 
This task is frequently encountered in sponsored recommendation domain where the RS searches for items user will most likely interact with. The difficulty of this task is due to the sparsity of interactions and high-dimensional item space~\cite{cohen2015house,kasandr_dataset}.
In this paper, we tackle these challenges by training for an additional dense objective, next item prediction task, that allows to learn useful representations of user state and items. Our model also learns interactions on the top of these representations using the state-action fusion mechanism. It shows improvement on predicting the clicked recommended item compared to navigation or click-only baselines.

Our setup borrows ideas from the reinforcement learning (RL) setup. In RL the recommendation system (agent) takes actions (recommends items) to users (environment) and receives a reward (business objective, e.g. click)~\cite{zheng2018drn,deep-rl:reco}. Similarly to RL setup, we also consider a RS as an agent acting upon the user to achieve its goals. We note two differences. First, we consider the production recommendation agent being a blackbox whereas in RL setting the goal is to optimize the production policy. Although directly optimizing the agent promises better performance, in practice, the operational constraints may not allow for it. Second, we learn against unsupervised target whereas in RL setting the agent optimizes for the long-term expected return. Next item prediction task is appealing because it does not depend on a particular business objective and it is widely used as a transfer task~\cite{mikolov2010recurrent,ranzato2014video}.

To summarize, our main contributions are as follows: (i) we extend the sequential recommendation setting by considering interaction with recommended items, (ii) we propose a RNN model that includes the recommendation action representation and state-action fusion mechanism, (iii) we compare scalable ways to fuse recommendation and user state on two real-world large-scale datasets and show improved performance on predicting the next item, in particular, predicting clicked recommended items. 

\section{Related Work}
\label{sec:relatedwork}
Previous works studied modeling of user interaction with recommended items in the context of click prediction task. We overview this research and compare them to our work in Section~\ref{sec:clickpred}.

Our proposed model builds on the top of conditional RNNs architectures that aim to capture complementary information in RNNs. We review main approaches in application to RS and video prediction in Section~\ref{sec:crnn}.

Our setup relates to the works in the domain of counterfactual (CF) learning. These works focus on batch learning from data produced by some logging policy. We discuss the relationship to our work and review the main approaches in application to RS in Section~\ref{sec:cf}.

\subsection{Click prediction}
\label{sec:clickpred}
In the click prediction task, the goal is given historical data about user activity and a list of suggested items, predict what items user is going to interact with. This task has frequently appeared in competitions, for example, Kaggle~\cite{outbrain:dataset} and RecSys~\cite{xing:dataset}. Recently,~\cite{kasandr_dataset} applied state-of-the-art recommender models to click prediction on a large-scale dataset. The winning approaches to this task typically include careful engineering of user and item features combined with strong classifiers, e.g. FFM~\cite{ffm_criteo}, GBDT~\cite{xing:winner,romov2015recsys} or DNN~\cite{cheng2016wide,guo2017deepfm}.
These works highlight the importance of constructing good representation of user and item. 
In this paper, we train for an additional next item prediction objective to automatically learn useful representations. We also build-in a feature interaction module, that we call \textit{state-action fusion}, that acts similar to a strong classifier by allowing for non-linear feature interactions. 

\subsection{Conditional Recurrent Neural Networks}
\label{sec:crnn}
\paragraph{Conditional RNNs for Recommendation} 
Hidasi et al.~\cite{rnn:balazs} first applied RNNs to the task of session-based recommendation. Authors showed that RNNs significantly outperforms the collaborative filtering baseline on predicting the next item in the user session on e-commerce website.
~\cite{smirnova2017contextual} improves on the top of~\cite{rnn:balazs} by leveraging contextual information such as type of user interaction and time difference between subsequent events. They studied two ways of introducing context in sequence models: (1) by modeling context-dependent input/output representation, where item representation is combined with context through multiplicative integration or concatenation, and (2) by modeling context-dependent dynamics, where context is used to parametrize the dynamics of the hidden state transition. In this work, employ similar conditioning mechanisms to model the recommendation action.

\paragraph{Conditional RNNs for Video Prediction} 
In the domain of vision-based reinforcement learning, previous works studied action-conditional video prediction problem. It consists of predicting the next frame in a video game based on the control actions taken at the current frame and previous frames.
Oh et al.~\cite{oh_et_al} proposed to incorporate action representation at the output layer of the RNN. In particular, they used multiplicative fusion of action representation and the hidden state to directly predict the next frame. Further we will refer to this mechanism as \textit{late state-action fusion}.
An extension to this work was proposed by Chiappa et al. in~\cite{chippa_et_al}. Authors argued that output layer fusion of~\cite{oh_et_al} influences state transitions only indirectly through the predictions. Therefore, they proposed to directly condition hidden dynamics on action using multiplicative interaction. In the following, we will refer to this alternative as \textit{early state-action fusion}.

\subsection{Learning from logged data} 
\label{sec:cf}
CF learning~\cite{swaminathan2015batch} differs from supervised learning in that the logged data only contains partial information limited to the actions taken by the logging policy. To account for the biased data collection, a popular approach is propensity weighting. It consists in re-weighting the loss terms by the probability of observing a particular training example, e.g. in recommendation setup, a user-item pair~\cite{reco:thorsten}. To avoid high variance problem, it is common to focus on non-sequential i.i.d setup. Another line of research in this domain takes the model-based approach that consists in defining a joint likelihood of the missing data model and the user feedback model~\cite{hernandez2014probabilistic}. Our work can be seen as an instance of model-based stateful CF learning since we implicitly define expectancies over the missing data by specifying the model of user state throughout the sequence and the impact of RS on it. 

Related but different from the problem we consider is training feedback loop issue that arises when the same RS acts as a logging policy and as a learning policy. 
To address this issue, authors in~\cite{li2010exploitation,Li:2010:CAP:1772690.1772758} take the perspective of exploration and exploitation trade-off. These works apply a variant of multi-armed bandit algorithm that learns to sequentially select items while optimizing for long-term expected reward. Differently, authors in~\cite{sinha2016deconvolving} approach the problem from model-based point of view. They consider a item-based RS where the observed feedback is a mixture of true feedback and recommended items. They propose an SVD-based method that deconvolves the effect of RS from the observation matrix. Compared to these works, we are interested in producing recommendations from logged data as opposite to uncovering the true ratings or intervening in production system.

\section{Proposed Approach}
\label{sec:proposedapproach}
As discussed in Section~\ref{sec:relatedwork}, click prediction and the next event prediction has been actively studied separately. In this paper, we propose a unified model of user behavior by considering user interactions with recommended items in addition to navigation history.

\subsection{Model}
\label{sec:proposedapproach:model}

\subsubsection{Setup}
\label{sec:proposedapproach:model:setup}

\begin{figure}
\begin{center}
\includegraphics[scale=0.3]{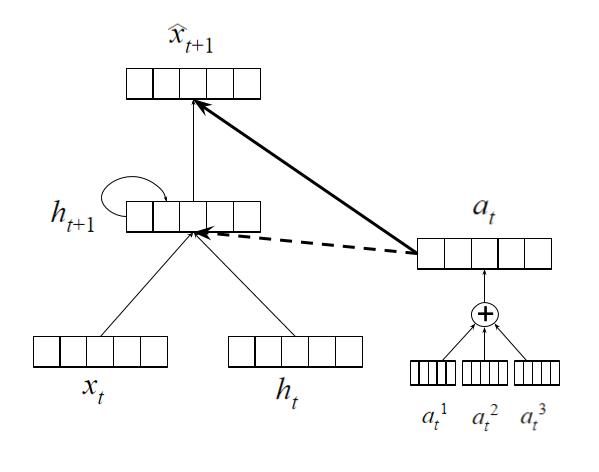}
\caption{Graphical overview of proposed model: $x_t$ is the visited item at time $t$, $a^i_t$ are the recommended items, $h_t$ is the hidden state of the RNN, the early state-action fusion is denoted by bold solid line, the late state-action fusion is is denoted by bold dotted line.}
\label{fig:ac-t2v}
\end{center}
\end{figure}

We are given sequence of observations $X=\left\{(x_t, a_t)\right\}, t=\overline{1,T}$, where $x_t \in R^{V_{x}}$ is one-hot encoded item id visited by the user at time step $t$, $a_t=[a^1_t;..;a^{k_t}_t], a^i_t \in R^{V_{x}}$ is concatenation of $k_t$ one-hot encoded recommended item ids $a^i_t$ at time step $t$. 

It is common to transform the original sparse item vector $x_t$ into dense item embedding using a linear mapping $\boldsymbol{x_t} = V^{embed}x_t$.

Our task consists in modeling the probability $p(x_{t+1}| a_t, x_{\leq t})$ of the next item $x_{t+1}$ given the recommended items $a_t$ and the history of user activity $x_{\leq t}$. 

\subsubsection{Recurrent architecture}
\label{sec:proposedapproach:model:recurrent}
We build our model on the top session-based RNN model proposed by Hidasi et al.\cite{rnn:balazs}:
\begin{eqnarray}
\label{eq:t2v}
\begin{aligned}
p(x_{t+1}| x_{\leq t}) & \propto f^{out}(h_{t+1}), \\
h_{t+1} &= \phi(\boldsymbol{x_t}, h_t),
\end{aligned}
\end{eqnarray}
where $f^{out}$ is the output layer of the RNN that returns a vector in original item id space, $\phi$, a is cell module, such as Gated Recurrent Unit (GRU) \cite{gru:cho} or Long Short Term Memory (LSTM) \cite{lstm:schmidhuber}, $h_t \in R^{k}$ is a state vector of the RNN, $k$ is the number of hidden state dimensions.

Based on previous works on action-conditional RNNs discussed in Section~\ref{sec:crnn}, we augment the model given by Eq.~\ref{eq:t2v} as follows:
\begin{eqnarray}
\label{eq:ac_t2v}
\begin{aligned}
p(x_{t+1}| a_t, x_{\leq t}) & \propto f^{out}(\tilde{h}_{t+1}), \\
\tilde{h}_{t+1} &= f^{late-fusion}(h_{t+1}, a_t), \\
h_{t+1} &= \phi(\boldsymbol{x_t}, \tilde{h}_t), \\
\tilde{h}_t &= f^{early-fusion}(h_t, a_t).
\end{aligned}
\end{eqnarray}
Figure~\ref{fig:ac-t2v} gives a graphical overview of the proposed model.

The difference with the baseline model from Eq.~\ref{eq:t2v} is that the next hidden state is computed using one of two fusion mechanisms. In the case of early state-action fusion $f^{early-fusion}$, the \textit{current} RNN hidden state is updated. In the case of late state-action fusion $f^{late-fusion}$, the \textit{next} RNN hidden state is updated. We describe the fusion mechanisms in Section~\ref{sec:proposedapproach:model:fusion}. 

Late fusion modifies the output layer of the RNN leading to a residual effect of the recommendation on the state of the RNN. In contrast, early fusion directly changes the state by integrating the action. Our particular choices of fusion mechanism are motivated by simplicity of implementation and scalability. We leave a broader study of fusion variants for future work.

The choice of the recommendation action representation $a_t$ is discussed in the next section. 

\subsubsection{Recommendation action representation}
\label{sec:proposedapproach:model:actionrepresentation}
Contrary to previous works on action-conditional video prediction, the RS action $a_t$ is high-dimensional and consists of a list of item ids with possibly varying size $k_t$.

We introduce embedding layer to construct a dense representation for each recommended item id $a^i_t$. For improved parameter sharing, we propose to reuse the linear projection $V^{embed}$ used to map the visited item: $\boldsymbol{a^i_t} = V^{embed} a^i_t$.

Varying size action representation being impractical, we obtain a fixed size action representation by averaging recommended item representations: $\boldsymbol{a_t} = \frac{1}{k_t}\sum_{i=1}^{k_t}{\boldsymbol{a^i_t}}$. We note that this representation is invariant to the order of items in the list.

We believe that recommendation action representation can be further improved by incorporating information about color, shape, position of the recommendation block. We plan to pursue this direction in future work.

\subsubsection{State-action fusion}
\label{sec:proposedapproach:model:fusion}
Similar to the multiplicative fusion of action-conditioned LSTM by Chiappa et al.~\cite{chippa_et_al} and action-conditional transformation by Oh et al.~\cite{oh_et_al}, we propose a parameter-efficient multiplicative integration of action representation $\boldsymbol{a_t}$ and RNN hidden state $h_t$:
\begin{align}
f^{early-fusion}(h_t, a_t) &= h_t \odot W^a \boldsymbol{a_t}, \label{eq:earlyfusion}\\
f^{late-fusion}(h_{t+1}, a_t) &= h_{t+1} \odot W^a \boldsymbol{a_t}, \label{eq:latefusion}
\end{align}
where $\odot$ denotes element-wise product, $W^a$ is the projection matrix between action space and RNN hidden state space.


\subsubsection{Optimization objective}
\label{sec:proposedapproach:model:optim}
We train the model by minimizing the negative log-likelihood of the training data:
\begin{eqnarray}
\label{eq:loss}
L = - \sum_{i=1}^N \sum_{t=1}^{T^{(i)}} \log p(x^{(i)}_{t+1}| a^{(i)}_t, x^{(i)}_{\leq t}),
\end{eqnarray}
where $N$ is the number of training sequences, $T^{(i)}$ is the length of $i$-th sequence.

\section{Experimental evaluation}
\label{sec:experiments}

\begin{table*}[!htbp]
\centering
\begin{tabular}{lrrrrrr} 
\toprule
\textbf{Model} & \multicolumn{3}{c}{\textbf{Outbrain}} & \multicolumn{3}{c}{\textbf{Internal dataset}}\\
& Global & View & Click & Global & View & Click \\
\toprule
\textit{Baselines}\\
Navigation RNN & 0.2943 $\pm 0.0004$ & 0.2995 $\pm 0.0007$ & 0.2098 $\pm 0.0040$ & 0.3558 $\pm 0.0004$ & 0.3635 $\pm 0.0006$ & 0.2101 $\pm 0.0040$\\
Clicks RNN & 0.0691 $\pm 0.0001$ & 0.0670 $\pm 0.0005$ & 0.1353 $\pm 0.0030$ & 0.1272 $\pm 0.0003$ & 0.1303 $\pm 0.0004$ & 0.1701 $\pm 0.0050$ \\
\midrule
\textit{Action-conditional RNNs}\\
Late Fusion & \textbf{0.3022} $\pm 0.0004$ & 0.3039 $\pm 0.0007$ & \textbf{0.2635} $\pm 0.0040$ & \textbf{0.3691} $\pm 0.0004$ & 0.3758 $\pm 0.0007$ & \textbf{0.2881} $\pm 0.0040$\\
Early Fusion & 0.2947 $\pm 0.0003$ & 0.2989 $\pm 0.0007$ & 0.2361 $\pm 0.0030$ & 0.3498 $\pm 0.0004$ & 0.3703 $\pm 0.0006$ & 0.2683 $\pm 0.0030$ \\
\bottomrule
\end{tabular}
\caption{Performance in terms of Precision@10 of action-conditional RNNs versus baselines. 'View' denotes navigational events, 'click' denotes clicked recommendation events. Best results are in bold. Uplift over the best baseline represents +40\% on recommendation events and +0.4\% globally on Criteo dataset. 95\% confidence intervals are obtained from 30 bootstraps.}
\label{tab:dataset-results}
\end{table*}

\subsection{Setup}
We evaluate the recommendation methods on the next item prediction task. We focus on evaluating the quality of top K recommendations. As evaluation metric, we use Precision at K (Precision@K) averaged over all events. Precision at K is the proportion of time the test item appears in the top K list of predicted items. This metric was used for evaluation in session-based recommendation~\cite{hrnn} and click prediction task~\cite{xing:dataset,kasandr_dataset}.

We compare our model against two baselines:
\begin{itemize}
\item \textit{Navigation RNN}: the RNN model defined in Eq.~\ref{eq:t2v} that models the user navigation but not the interaction with recommended items,
\item \textit{Clicks RNN}: the action-conditional RNN model defined in Eq.~\ref{eq:ac_t2v} with late fusion but only trained to predict clicked recommended items; this baseline addresses the click prediction task with user state optimized to predict clicked item.
\end{itemize}

\subsection{Action-conditional RNNs}
Based on the recurrent architecture described in Section~\ref{sec:proposedapproach:model}, we experiment with the following architectures:
\begin{itemize}
	\item \textit{Early Fusion}: Multiplicative integration of hidden state and action at RNN state transition, given by Eq.~\ref{eq:earlyfusion},
    \item \textit{Late Fusion}: Multiplicative integration of hidden state and action at RNN output layer, given by Eq.~\ref{eq:latefusion}.
\end{itemize}

\subsubsection{Hyper-parameters}
We fixed the size of item embedding vector and the size of RNN hidden state to 40. For optimization of the loss function in Eq.~\ref{eq:loss}, we use the Adam algorithm with squared root decay of learning rate from 0.01 to 0.001. For all models, the batch size was set 64 and number of training iterations to 10,000, the cell function $\phi$ is the GRU cell.


\subsection{Datasets}
We experimented on two datasets. The first dataset is publicly available Outbrain dataset released for Kaggle competition~\cite{outbrain:dataset}. It contains a sample of user viewing activity on multiple publisher sites in the United States over the period of 14-28 Jun 2016. The second dataset is proprietary dataset that contains user browsing activity on multiple e-commerce websites over the period of Oct-Dec 2017. We refer to this dataset as Internal dataset. 

In both datasets, the user activity sequence includes the sets of recommendations served to a specific user where at least one recommendation was clicked. We note that the percentage of events with recommendation interaction is 6.8\% for Outbrain dataset and 1.5\% for Internal dataset. These rates are typical for sponsored recommendation industry.

In terms of pre-processing, we replaced product ids with less that 10 occurrences with a single id. This resulted in 225K distinct items for Outbrain dataset and 210K for Internal dataset. We kept 40 latest events in each sequence. We capped the maximum number of products in recommendation set to 5. 

We performed training/validation split by randomly selecting user sequences. We obtained a split of 4M/1.5M for Outbrain dataset and 2M/500K for Internal dataset. 

\subsection{Results}
Table \ref{tab:dataset-results} summarizes the performance of action-conditional RNNs versus baseline in terms of Precision@10. 
Given that the majority events in the datasets are navigation events without interaction with recommendation, we additionally show the performance break down on navigation events (denoted 'view') and on events with clicked recommendation (denoted 'click'). 

The clicks RNN baseline performs significantly worse than all action-conditional RNNs that are additionally trained to predict the next item. We conclude that this objective leads to transferable representations to click prediction task.

We observe that Navigation RNN baseline has substantially lower precision at clicked recommendation events than at view events. Thus, it is harder to predict the user transition at recommendation events from user navigation alone.

We find that action-conditional RNN with late fusion effectively incorporates recommended items information. It outperforms the baselines by improving globally and on recommendation events. The uplift over the best baseline represents +0.4\% of Precision@10 globally and +40\% of Precision@10 on clicked recommendation events for Criteo dataset and +25\% for Outbrain dataset. We note that this uplift is significant both statistically and practically for sponsored recommendation industry. As expected, the difference in performance on view events is not statistically significant since late fusion mechanism only introduces an indirect influence on user state through output layer.

In contrast, the early fusion degrades the performance overall. We hypothesize that the recommendation is having a residual effect on the user state as opposite to taking actions in video games. Therefore, the indirect state-action fusion at output layer is more suitable in recommendation case.


\section{Conclusion}
\label{sec:conclusion}
In this paper, we introduced a more general setup of sequential recommendation with interaction with recommended items. We proposed as a solution the action-conditional RNN architectures and compared several scalable state-action fusion mechanisms. From our experiments, we conclude that late fusion mechanism is more suitable in recommendation case. In particular, action-conditional RNN with late fusion shows both practically and statistically significant uplift (+40\%) on sponsored recommendation events.

As future work, we intend to investigate other variants to fuse state and action suitable for recommendation scenario. Improving action representation is another direction we plan to study.

\bibliographystyle{ACM-Reference-Format}
\bibliography{literature}

\end{document}